\documentstyle[12pt]{article}

\newcommand{\Od}{{\cal O}}

\newcommand{\Tr}{\mbox{Tr}}
\newcommand{\im}{\mbox{Im}}

\begin{document}

\title{Non-local gravitational effective action and
particle production\footnote{Contribution to the Proceedings of the
International Seminar on Mathematical Cosmology, Potsdam (Germany),
March 30 - April 4, 1998}}

\author{A. Dobado\\{\it Departamento de F\'{\i}sica Te\'orica} \\
{\it Universidad Complutense de Madrid}\\
 {\it 28040 Madrid, Spain}\\ and \\
A.L. Maroto\\
{\it Departamento de F\'{\i}sica Te\'orica} \\
{\it Universidad Aut\'onoma de Madrid}\\
{\it 28049 Madrid, Spain}}


\maketitle\begin{abstract}
We study the effective action for 
gravity obtained
after the integration of scalar matter fields, using
the local momentum representation based on the Riemann normal
coordinates expansion. By considering this expansion around 
different space-time points,
we also compute  the non-local terms together with the more usual 
divergent ones. We discuss the applicability of our results to 
the calculation of 
particle production rates in cosmological backgrounds 
and compare this method with the traditional
Bogolyubov transformations.
\end{abstract}

\section{Introduction}
One of the most interesting aspects of quantum field theory in the 
presence of a external classical background is the possibility of
particle creation. When a external field $J$ is present, the vacuum
state (with zero particle number) can become unstable and eventually 
decay by particle emission. In other words, if the system is prepared in
the vacuum state in the remote past $\vert 0, in\rangle$, the probability
that the system remains in the same state in the far future will 
be in general
different from one: 
$\vert\langle 0, in\vert 0, out \rangle_J\vert ^2\neq 1$.

The first works about this subject can be traced
back to the early fifties in the context of quantum electrodynamics (QED)
\cite{Schwinger}. In particular, it was shown that  intense 
electric fields could give rise to the creation of electron-positron pairs.
The corresponding particle creation rates were obtained by means of 
effective action (EA) techniques. Within this formalism the vacuum 
persistence
amplitude $\langle 0, in\vert 0, out \rangle_J$ is given by means of 
the path integral. 
In the simple case of a real
scalar field in a external gravitational background, the EA is given by:
\vspace{-0.3cm}
\begin{eqnarray}
\langle 0, in\vert 0, out \rangle_{g_{\mu\nu}}&=& 
e^{iW[g_{\mu\nu}]}
=\int [d\phi]e^{iS[g_{\mu\nu},\phi]}\nonumber \\&=&
  \int [d\phi]e^{-\frac{i}{2}\int d^4x\sqrt{g}\phi
(\Box+m^2+\xi R -i\epsilon)\phi}=
(\det O)^{-1/2}
\label{eaes}
\end{eqnarray}
where  
$O_{xy}(m^2)=(-\Box_y-m^2-\xi R(y) +i\epsilon)\delta^0(x,y)$ 
with $\delta^0(x,y)$ the covariant Dirac delta. As usual we introduce
the non-minimal coupling $\xi R\phi^2$ in order to include the case with
conformal invariance $\xi=1/6$.
As we see from this expression, the functional integration is performed
only on the scalar fields, i.e, they will appear inside the loops, whereas
the gravitational field $g_{\mu\nu}$ is considered as classical and 
hence it will only appear in the external lines. As a consequence,
$W[g_{\mu\nu}]$ is the generating functional of Green 
functions with external
gravitational lines and loops only of matter fields. Another interesting
point is that as far as the classical action is quadratic in the scalar
fields, the one loop calculation is exact, i.e there is no higher loops
contribution to the EA.

It is now relatively easy to compute the pair production probability 
$P_{EA}$
that
according to the above discussion will be nothing but:
$P_{EA}=1-\vert\langle 0, in\vert 0, out \rangle_{g_{\mu\nu}}\vert ^2$. 
From 
eq.\ref{eaes} assuming $W[g_{\mu\nu}]$ to be small, we obtain:
\begin{eqnarray}
P_{EA}=1-e^{-2\im W[g_{\mu\nu}]}\simeq 2\im W[g_{\mu\nu}]
\end{eqnarray}
Thus the imaginary part of the effective action 
provides the particle creation probabilities.

Since the first developments of quantum field theory in curved space-time,
the topic of particle creation by  gravitational fields has received 
much attention. The pioneering works were due to Parker \cite{Parker} who
studied particle production in Robertson-Walker geometries by means of the
mode-mixing Bogolyubov method \cite{Birrell}. Some other important works 
are those of
Zel'dovich and Starobinsky \cite{Zeldovich,Hartle} that extended this 
technique to
homogeneous but anisotropic cosmological space-times. They show 
that particle
production provides an effective mechanism for anisotropy damping in the
early universe. But probably it is Hawking radiation the 
best known example of particle production from a gravitational field.
Most of these calculation were done by means of Bogolyubov method. However
in this work we present an alternative technique for the calculation 
of particle
creation based on the gravitational EA. 

The gravitational effective action eq.\ref{eaes} is in general a 
non-local functional
on the metric tensor and the connection. 
Although the divergent 
local terms have been extensively considered
in the literature \cite{Birrell}, the finite non-local contributions have
been studied only more recently \cite{Hartle,Vilkovisky,Avramidi,DOMA1}. 
The non-local terms
are responsible for the imaginary part and therefore also for
the particle creation. The calculation of the EA is in general 
very difficult
for an arbitrary space-time geometry and accordingly we have to rely on
some kind of approximation. The most traditional one is the well-known
perturbation theory in which the metric tensor is splitted as follows:
$g_{\mu\nu}=\eta_{\mu\nu}+\kappa h_{\mu\nu}$. The EA is then
expanded in powers of the coupling $\kappa$. This method however is not 
generally covariant.
We will present an alternative (covariant) approximate method in which 
the EA is 
expanded in powers of curvature tensors. The curvature expansion will
be generated in turn from the Riemann normal coordinates expansion.
This kind of curvature expansions  was first considered by Barvinsky and 
Vilkovisky \cite{Vilkovisky}.

\section{Non-local gravitational effective action}

Let us consider a smooth manifold in which geodesics do
not intersect so that we can perform the well-known \cite{Petrov} 
Riemann normal coordinates expansion for the metric tensor:
\begin{eqnarray}
g_{\mu\nu}(y)&=&\eta_{\mu\nu}+\frac{1}{3}
R_{\mu\alpha\nu\beta}(y_0)y^\alpha y^\beta-
\frac{1}{6}R_{\mu\alpha\nu\beta ;\gamma}(y_0)
y^\alpha y^\beta y^\gamma \nonumber \\
&+&
\left[\frac{1}{20}R_{\mu\alpha\nu\beta ;\gamma\delta}(y_0)
+\frac{2}{45}R_{\alpha\mu\beta\lambda}(y_0)
R^{\lambda}_{\;\gamma\nu\delta}(y_0)\right]
y^\alpha y^\beta y^\gamma y^\delta +\Od (\partial^5)
\label{normal}
\end{eqnarray}
where $y_0$ is the origin of normal coordinates. When substituting this
expansion in eq.\ref{eaes} we will generate the curvature expansion we
are looking for.  The operator $O_{xy}$ defined above can then be 
splitted in a free part:
\begin{eqnarray}
A_{xy}(m^2)=(-\Box_0^y-m^2 +i\epsilon)\delta^0(x,y)
\end{eqnarray}
with $\Box_0^y=\eta^{\mu\nu}\partial_\mu^y \partial_\nu^y$, and
an interaction term that contains all the curvature
dependence:
\begin{eqnarray}
B_{xy}&=&\left[-\frac{2}{3}R^\lambda_{\;\;\rho}(y_0)
y^\rho\partial_\lambda^y+\frac{1}{3}R^{\mu\;\;\nu}
_{\;\;\epsilon\;\;\beta}(y_0)y^\epsilon y^\beta 
\partial_\mu^y\partial_\nu^y
-\xi R(y_0)\right. \nonumber \\
&-&\left(\frac{1}{20}R_{\alpha\;\; ; \beta\gamma}^{\;\;\nu}(y_0)
+\frac{1}{20}R_{\alpha\;\; ; \gamma\beta}^{\;\;\nu}(y_0)
-\frac{1}{20}R^{\mu\;\;\nu}_{\;\;\alpha\;\;\beta;\mu\gamma}(y_0)
-\frac{1}{20}R^{\mu\;\;\nu}_{\;\;\alpha\;\;\beta;\gamma\mu}
(y_0)\right.\nonumber\\
 &-&\frac{8}{45}R_{\alpha\lambda}(y_0)
R^{\lambda\;\;\nu}_{\;\;\beta\;\;\gamma}(y_0)
+\frac{1}{15}R_{\alpha\;\;\beta\lambda}^{\;\;\mu}(y_0)
R^{\lambda\;\;\nu}_{\;\;\mu\;\;\gamma}(y_0)
+\frac{4}{45}R_{\alpha\;\;\beta\lambda}^{\;\;\mu}(y_0)
R^{\lambda\;\;\nu}_{\;\;\gamma\;\;\mu}(y_0)
\nonumber \\
&+&\left.\frac{1}{40}R_{\alpha\beta;\gamma}^{\;\;\;\;\;\;\;\nu}(y_0)
+\frac{1}{40}R_{\alpha\beta;\;\;\gamma}^{\;\;\;\;\;\nu}(y_0)\right)
y^\alpha y^\beta y^\gamma \partial_\nu^y 
-\left(-\frac{1}{20}
R^{\mu\;\;\nu}_{\;\;\rho\;\;\epsilon;\delta\kappa}(y_0)\right.
\nonumber \\
&+&\left.\left.
\frac{1}{15}R_{\rho\;\;\epsilon\lambda}^{\;\;\mu}(y_0)
R^{\lambda\;\;\nu}_{\;\;\delta\;\;\kappa}(y_0)
\right) y^\rho y^\epsilon y^\delta y^\kappa 
\partial_\mu^y\partial_\nu^y -\xi\frac{1}{2}
R_{;\alpha\beta}(y_0)y^\alpha y^\beta\right]\delta^0(x,y)+... \nonumber \\
\label{B}
\end{eqnarray}
Accordingly we can write:
\begin{eqnarray}
W[g_{\mu\nu}]=\frac{i}{2}\Tr\log O(m^2)
=\frac{i}{2}\Tr \log(A+B)
\end{eqnarray}
It is then possible to expand $W$ in a power series in the 
interaction operator $B$. With that purpose we first take the derivative
of the EA with respect to the mass parameter, in order to avoid the 
problem of series expansions of logarithms of operators \cite{DOMA1}:
\begin{eqnarray}
\frac{d}{dm^2}W[g_{\mu\nu}]&=&-\frac{i}{2}\Tr \frac{1}{O(m^2)}=
-\frac{i}{2}\Tr((A+B)^{-1})\nonumber \\
&=&-\frac{i}{2}\Tr(A^{-1}-A^{-1}BA^{-1}+A^{-1}BA^{-1}BA^{-1}-...)
\label{invdes}
\end{eqnarray}
As far as the $B$ operator is at least linear in the curvatures and
we are only interested in the expansion up to order quadratic
in the curvatures, we will only have to keep the first three terms.
We now face the main problem of normal coordinates expansion. As we can
see from eq.\ref{B}, all the curvature tensor are evaluated at the same
point $y_0$, and therefore eq.\ref{invdes} can hardly gives rise to a
non-local expression. However, it is possible to circumvent this problem 
by means of a point splitting procedure in the quadratic terms. Using
again the Riemann normal coordinate expansion we can write:
\begin{eqnarray}
{\cal R}(y_0){\cal R}(y_0)={\cal R}(y_0){\cal R}(y)
+({\cal R}(y_0)\nabla{\cal R}(y_0)y+
{\cal R}(y_0)\nabla^2{\cal R}(y_0)yy+...)\nonumber \\
+({\cal R}(y_0){\cal R}(y_0){\cal R}(y_0)yy+
{\cal R}(y_0)\nabla{\cal R}(y_0){\cal R}(y_0)yyy+...)+...
\label{pointsplit}
\end{eqnarray}
where ${\cal R}$ collectively denotes the scalar curvature or the
Riemann and Ricci tensors. There are several ways of performing the
splitting (according to the chosen $y$ point), however they all will
differ in higher order terms in derivatives. 

Evaluating the functional traces in eq.\ref{invdes} by using dimensional
regularization, we obtain the final result for the gravitational EA
up to $\Od({\cal R}^2)$ for a real scalar field. We show 
the masless limit \cite{DOMA1}:
\begin{eqnarray}
W[g_{\mu\nu}]&=&\int d^4 x \frac{\sqrt{g}}{32\pi^2}
\left(\frac{1}{180}\left(R^{\mu\nu\lambda\rho}(x)\Gamma(\Box)
R_{\mu\nu\lambda\rho}(x)-R^{\mu\nu}(x)\Gamma(\Box)R_{\mu\nu}(x)\right)
\right.  \nonumber\\
&+& \left. 
\frac{1}{2}\left(\frac{1}{6}-\xi\right)^2 R(x)\Gamma(\Box)R(x)\right)
+\Od ({\cal R}^3)
\label{nolocal}
\end{eqnarray}
with $\Gamma(\Box)=N_\epsilon-\log(\Box/\mu^2)$ and 
$N_\epsilon=2/\epsilon +\log 4\pi -\gamma$
the usual way of parametrizing the poles in dimensional regularization.
The action of the $\Gamma(\Box)$ operators should be understood through
the corresponding Fourier transform in normal coordinates, i.e:
\begin{eqnarray}
\log\left(\frac{\Box}{\mu^2}\right){\cal R}(y_0)=
\int d^4x \frac{d^4p}{(2\pi)^4}e^{ip x}
\log\left(\frac{-p^2-i\epsilon}{\mu^2}\right)
{\cal R}(x)
\label{act}
\end{eqnarray}
 We see that the massless limit is regular \cite{DOMA1} and in this
case, the above expression for the EA is exact up to quadratic order.

\section{Particle production in cosmological backgrounds}

In this section we are interested in calculating the imaginary part
of the gravitational EA in eq.\ref{nolocal} that, as shown in the 
introduction, provides
us with the particle production probability. 

Let us consider a general Bianchi I type metric:
\begin{eqnarray}
ds^2=a^3(\tau)d\tau^2-a^2(\tau)g_{ij}(\tau)dx^idx^j
\end{eqnarray}
As far as
this metric is homogeneous, it is possible to write the different curvature
tensor in such a way that they only depend on the time coordinate, and 
therefore:
\begin{eqnarray}
&\im & \int d^4x \frac{d^4p}{(2\pi)^4}e^{ip x}{\cal R}(y_0)\log 
\left(\frac{-p^2-i\epsilon}{\mu^2}\right){\cal R}(x)\\
&=&
\im \int dx^0 \frac{dp_0}{(2\pi)}e^{ip_0 x^0}
{\cal R}(y_0)\log\left(\frac{-p_0^2-i\epsilon}{\mu^2}\right){\cal R}(x^0)
=-\pi{\cal R}(y_0){\cal R}(y_0)\nonumber
\label{frwim}
\end{eqnarray}
Using this result it is possible to write the imaginary part as a 
purely local 
expression:
\begin{eqnarray}
\im \; W[g_{\mu\nu}]&=&\int d^4x \frac{\sqrt{g}}{32\pi}\left(
\frac{1}{120}C_{\mu\nu\rho\sigma}
C^{\mu\nu\rho\sigma}+\left(\frac{1}{6}-\xi\right)^2 \frac{R^2}{2}\right)
+\Od ({\cal R}^3)
\label{imbi}
\end{eqnarray}
Here $C_{\mu\nu\rho\sigma}$ is the Weyl tensor and we have discarded a
Gauss-Bonnet term. In order to do so, it is necessary that space-time 
is asymptotically flat or at least that asymptotically 
($\tau\rightarrow \pm \infty$), 
the scale factor behaves as: 
$a(\tau)\sim \vert \tau\vert^\alpha$ with $\alpha >-1/2$. 
The integrand can be readily interpreted as the particle 
production probability
per unit time and volume. Therefore, unlike Bogolyubov method that only
provides the spectrum of the particle produced, the EA method allows 
us to know the
instantaneous particle production rate. This rate will give 
rise to the same
total amount of particle as the Bogolyubov method in contrast with other
methods such as Hamiltonian diagonalization (see \cite{Birrell} 
and references therein)
which produce vastly
more particles. Eq.\ref{imbi} is just the general form of 
the Zel'dovich and 
Starobinski \cite{Zeldovich} result. When the metric tensor reduces 
to the Friedmann-Robertson-Walker (FRW) 
form (with vanishing Weyl tensor), the probability will only depend on
the scalar curvature. It is then easy to see that in the case of 
conformal invariance ($\xi=1/6$) there is no particle production, 
as well as in a
radiation dominated universe for which $R=0$ \cite{Parker}.

It is possible to compare (for specific FRW models with $\xi=0$ and complex
scalar field), 
the particle production probabilities per unit volume 
($p_{EA}$ and $p_{BOG}$) derived with both methods:
\begin{eqnarray}
p_{BOG}\simeq \int \frac{d^3 k}{(2\pi)^3} \log (1+ <N_k>)\;\;\mbox{and}\;
p_{EA}\simeq \frac{1}{576\pi}\int d\tau a^6(\tau)R^2(\tau)
\label{earw}
\end{eqnarray} 
As an example let us take the model considered in \cite{schafer} with 
$a^4(\tau)=A^2\tau^2+B^2$, 
where $A$ and $B$ are arbitrary constants.
The number density of created pairs in the $k$ mode is given by 
the Bogolyubov method $<N_k>=\exp(-\frac{\pi B^2 k}{A})$. 
From eq.\ref{earw} we extract the following analytical results:
\begin{eqnarray}
p_{BOG}=p_{EA}=\frac{7 A^3}{360 B^6\pi}
\end{eqnarray}
i.e. both methods give the same probabilities.
Comparison with other models can be found in \cite{DOMA2}. Finally it is
worth mentioning that the EA method  also provides the 
 particles spectra \cite{DOMA2} and could be 
useful in those contexts in which Bogolyubov
transformations have been traditionally applied; such 
as graviton production or 
the generation
of anisotropies in inflationary cosmological models \cite{bran}.

{\bf Acknowledgements:} This work has been partially 
supported by Ministerio de
Educaci\'on y Ciencia (Spain) CICYT (AEN96-1634).




\newpage
\begin{thebibliography}{99}

\bibitem{Schwinger} J. Schwinger, {\it Phys. Rev} {\bf 82}, 664 (1951)
\bibitem{Parker} L. Parker, {\it Phys. Rev. Lett.} {\bf 21}, 562 (1968);
{\it Phys. Rev.} {\bf 183}, 1057 (1969)
\bibitem{Birrell} N.D. Birrell and P.C.W.
Davies {\it Quantum Fields in Curved Space}, Cambridge University Press 
(1982)
\bibitem{Zeldovich} Y.B. Zel'dovich and 
A.A. Starobinski, {\it Pis'ma Zh. Eksp. Teor. Fiz}
 {\bf 26}, 
373 (1977) ({\it JETP Lett.} {\bf 26}, 252 (1977))
\bibitem{Hartle} J.B. Hartle and B.L. Hu, {\it Phys. Rev.} {\bf D20},
1772 (1979)
\bibitem{Vilkovisky} A.O. Barvinsky and 
G.A. Vilkovisky, {\it Nucl. Phys.} {\bf B282} (1987),
163 (1987); {\bf B333}, 512 (1990)\\
A.O. Barvinsky, Yu.V. Gusev, 
G.A. Vilkovisky and V.V. Zhytnikov, {\it Nucl. Phys.} {\bf B439} 
(1995)
\bibitem{Avramidi} I.G. Avramidi, {\it Nucl. Phys.} {\bf B355}, 712 (1991)
\bibitem{DOMA1} A. Dobado and A.L. Maroto, preprint hep-th/9712198 
\bibitem{Petrov} A.Z. Petrov, {\it Einstein Spaces}, Pergamon, Oxford (1969)
\bibitem{schafer} G. Sch\"afer, {\it J. Phys.} {A12}, 2437 (1979)
\bibitem{DOMA2} A. Dobado and A.L. Maroto, preprint gr-qc/9803076
\bibitem{bran}V.F. Mukhanov, H.A. Feldman and R.H.
Brandenberger, {\it Phys. Rep.} {\bf 215}, 203 (1992);
Y. Shtanov, J. Traschen and R. Branderberger, {\it Phys. Rev.} {\bf D51},
5438 (1995)
\end{thebibliography}
\end{document}